\documentclass[twocolumn,amsmath,amssymb,prb]{revtex4}

\usepackage{graphicx} 
\usepackage{dcolumn}  
\usepackage{bm}       

\begin{document}


\title{Strain-induced insulator state in La$_{0.7}$Sr$_{0.3}$CoO$_{3}$}

\author{A.~D. Rata$^{1}$, A. Herklotz$^{1}$, K. Nenkov$^{1,2}$, L. Schultz$^{1}$, and K.
D\"{o}rr$^{1}$}

\affiliation{$^{1}$IFW Dresden, Institute for Metallic Materials,
Helmholtzstra$\ss$e 20, 01069 Dresden, Germany \linebreak
$^{2}$International Laboratory of High Magnetic Fields and Low
Temperatures, PL-53421 Wroclaw, Poland}

\date{\today}

\begin{abstract}
We report on the observation of a strain-induced insulator state
in ferromagnetic La$_{0.7}$Sr$_{0.3}$CoO$_{3}$ films. Tensile
strain above $1\%$ is found to enhance the resistivity by several
orders of magnitude. Reversible strain of $0.15\%$ applied using a
piezoelectric substrate triggers huge resistance modulations,
including a change by a factor of $10$ in the paramagnetic regime
at $300$~K. However, below the ferromagnetic ordering temperature,
the magnetization data indicate weak dependence on strain for the
spin state of the Co ions. We interpret the changes observed in
the transport properties in terms of a strain-induced splitting of
the Co $e_{g}$ levels and reduced double exchange, combined with a
percolation-type conduction in an electronic cluster state.

\end{abstract}

\pacs{...}

\maketitle



The electronic properties of some transition metal (TM) oxide
compounds are highly sensitive to external parameters such as
magnetic and electric fields, pressure, and chemical doping. One
prominent example are the manganese perovskites
R$_{1-x}$A$_{x}$MnO$_3$ showing the colossal magnetoresistance
(CMR) phenomenon, where R~=~La or a rare earth metal and
A~=~non-trivalent metal \cite{tokura_rev,dagotto,kathrin_rev}. The
width ($W$) of the TM $3d$ bands is typically low in these
materials and, thus, electronic transport is susceptible to
localization effects. $W$ can be modified by changing the
TM-oxygen bond lengths and/or the TM-O-TM bond angles. Hence,
hydrostatic pressure as well as epitaxial strain in thin films may
affect strongly the electronic structure and transport properties.
Beside the necessity to clarify the influence of strain for
applications of TM oxides in microelectronics, controlled strain
may provide an efficient tool to tune their electronic properties.

Recently, a strong sensitivity to pressure has been revealed for
cobaltites, e.g., in La$_{x}$Sr$_{1-x}$CoO$_3$
\cite{Lengsdorf,Lengsdorf2,Fita}. The conductivity of
La$_{0.82}$Sr$_{0.18}$CoO$_3$ single crystals was found to
decrease under pressure, although intuitively the width of the
conduction bands is expected to increase with pressure. A
pressure-induced insulator state has been observed and attributed
to a pressure-induced transition to a low-spin state for the
Co$^{3+}$ ions \cite{Lengsdorf,Lengsdorf2}. Perovskite-type
cobaltites possess thus an additional degree of freedom: the
Co$^{3+}$ $d^{6}$ and Co$^{4+}$ $d^{5}$ ions may display various
spin states. This is due to a delicate balance between the
crystal-field splitting $\Delta_{\mathrm{CF}}$ and the intraatomic
Hund exchange $J_{\mathrm{H}}$ \cite{raccah}. Since
$\Delta_{\mathrm{CF}}$ is very sensitive to the variation of the
Co-O bond length \cite{Sherman,Korotin}, structural changes may be
used to modify the ratio between $\Delta_{\mathrm{CF}}$ and
$J_{\mathrm{H}}$.

The ground state of the parent compound LaCoO$_3$ is nonmagnetic
with a low-spin (LS,~$S\!=\!0$) configuration, which can be
thermally excited either to an intermediate-spin (IS,~$S\!=\!1$)
or a high-spin (HS,~$S\!=\!2$) state above $T\!\sim\!100$~K
\cite{raccah,Korotin}. By doping with Sr$^{2+}$, the material
turns into a ferromagnetic cluster-glass for $x\!>\!0.18$,
following a spin-glass-like region between $x\!=\!0$ and
$x\!=\!0.18$ \cite{Itoh,goodenough95}. In the simplest picture,
doping is associated with the creation of Co$^{4+}$ ions and the
double exchange (DE) between Co$^{4+}$ and the adjacent IS
Co$^{3+}$ induces ferromagnetic couplings
\cite{goodenough95,Louca03}. The electrical conductivity increases
with $x$ and for $x\!\geq\!0.20$ the system is metallic
\cite{Itoh,goodenough95,Caciuffo,Lorenz}.

\begin{figure}[!b]
\includegraphics[width=7cm]{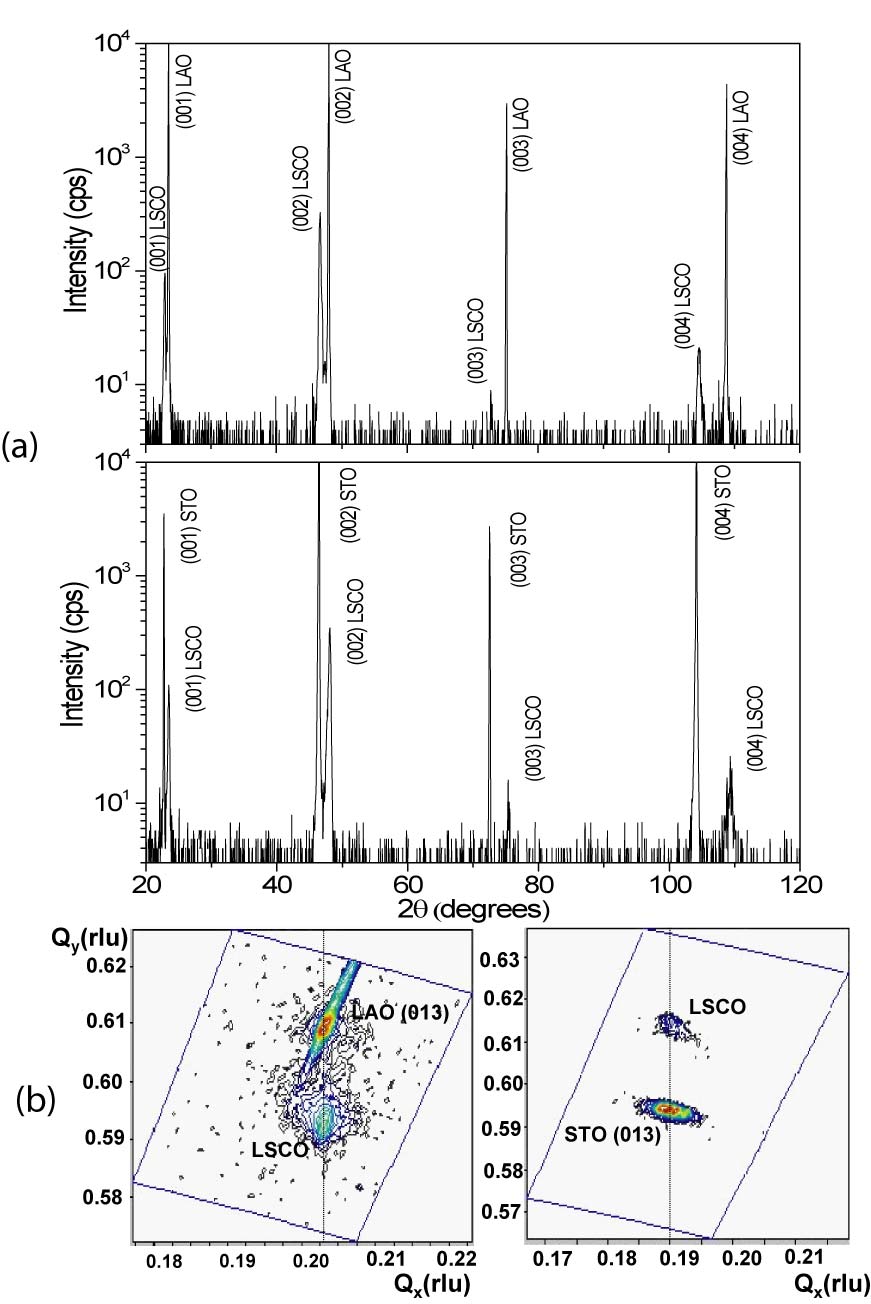}
\caption{1a) $\Theta-2\Theta$ X-ray diffractograms of $60$~nm
thick LSCO films, grown on  LAO (top) and STO (bottom) substrates.
1b) XRD reciprocal space map around the (013) reflection for LSCO
films on LAO (left) and STO (right).}
\end{figure}

Here, we investigate the influence of epitaxial strain on the
electronic transport and magnetization of
La$_{0.7}$Sr$_{0.3}$CoO$_{3}$ films epitaxially grown on various
substrates. The strain drastically modifies the conductivity of
our films, by several orders of magnitude. An insulator-type
behavior is observed in films grown under tensile strain, whereas
compressed films show bulk-like, metallic properties.
Nevertheless, the magnetization data reveal a much weaker response
to the strain, with rather small changes of the ordered magnetic
moments and ordering temperature $T_C$. The dependence on strain
of the resistance and magnetization of the
La$_{0.7}$Sr$_{0.3}$CoO$_{3}$ films was also measured by dynamic
strain experiments, using a pseudocubic piezoelectric substrate.
Once again, an extreme effect of strain on the transport is found,
while only minor changes of the magnetization were observed. We
discuss the experimental results in terms of a strain-induced
splitting of the Co $e_{g}$ levels and reduced DE, combined with a
percolation-type conduction in an electronic cluster state.

\begin{table}[t]
\caption{In-plane $a$ and out-of-plane $c$ lattice parameters of
the LSCO films, their tetragonal distortion $t$ defined as
$2(a\!-\!c)/(a\!+\!c)$, and the Curie temperature $T_\mathrm{C}$.
The lattice parameter ($a_{sub}$) of each substrate is also
given.}
\begin{ruledtabular}
\begin{tabular}{lccccc}
LSCO films     &$a_{sub}$ (\AA )
                           &$c$ (\AA )
                                        &$a$ (\AA )
                                                     &$t$
                                                                 &$T_\mathrm{C}$ (K)  \\
(60 nm)   \\
\colrule
\\
LSCO/LAO       &3.78       &3.87        &3.78        &$-2.3\%$   &194        \\
LSCO/STO       &3.905      &3.78        &3.90        &$+3.1\%$    &195        \\
LSCO/PMN-PT    &4.02       &3.79        &3.86        &$+1.8\%$    & 200          \\
\end{tabular}
\end{ruledtabular}
\end{table}

La$_{0.7}$Sr$_{0.3}$CoO$_{3}$ (LSCO) films were grown on
single-crystalline substrates of SrTiO$_3$(001)~(STO),
LaAlO$_3$(001)~(LAO), and
Pb(Mg$_{1/3}$Nb$_{2/3}$)$_{0.72}$Ti$_{0.28}$O$_3$(001)~(PMN-PT) by
pulse laser deposition (KrF $248$~nm) from a stoichiometric
target. The deposition temperature and the oxygen background
pressure were $650^{\circ}$~C and $3.5$x$10^{-1}$~mbar,
respectively. After deposition, the films were cooled down in
oxygen atmosphere of $600$~mbar. The deposition rate was
calibrated by measuring the film thickness by X-ray reflectivity.
X-ray diffraction (XRD) measurements for structural
characterization were carried out in a Philips X'Pert MRD
diffractometer using Cu K$\alpha$ radiation. The surface
microstructure of the films was probed by Atomic Force Microscopy
(AFM). The magnetization $M$ was measured in a SQUID magnetometer
in both field-cooled (FC) and zero-field cooled (ZFC) modes.
Electrical transport measurements using a standard four-point
technique were performed either with a Quantum Design PPMS system
or a split-coil magnet. In recent work \cite{thiele07,ramesh07} we
have proposed the reversible control of the biaxial in-plane
strain in films grown epitaxially on piezoelectric PMN-PT(001).
The substrate strain is controlled by applying an electrical
voltage between the conducting film and a bottom electrode on the
opposite face of the substrate (see inset in Fig.~4a). The
employed PMN-PT has a rhombohedral lattice structure similar to
that of LAO, apart from the larger lattice parameter of $4.02$ \AA
. If the LSCO films had too high resistance for the proper
function as an electrode, Pt electrodes were deposited on top,
which allows a two-point measurement of the LSCO resistance.

First, structural results for LSCO films grown epitaxially in
three distinct strain states, induced by varying the substrate
lattice parameter, will be discussed. In Fig.~1a we show the wide
angle $\Theta\!-\!2\Theta$ diffractograms of $60$~nm thick films
grown on LAO (top) and STO (bottom) substrates. The films display
clear (00l) reflections of the pseudocubic perovskite structure,
which are substantially shifted by the induced strain. The LSCO
film grown on LAO (STO) is under in-plane compressive (tensile)
strain. No impurity phase or deviating orientations were detected.
In Fig.~1b we compare the XRD reciprocal space maps around the
non-specular (013) reflection of LSCO films grown on LAO and STO
substrates. The intensity scale in the figure is logarithmic. The
horizontal and vertical axes are Q-vectors parallel
($Q_{\mathrm{x}}$) and perpendicular ($Q_{\mathrm{y}}$) to the
surface plane, respectively. In each case, the XRD peaks of the
LSCO film and the substrate, either LAO or STO, are observed at
the same $Q_{\mathrm{x}}$ values, proving that the growth is
coherent. The derived pseudocubic in-plane lattice parameters of
the films are $3.78$~$\mathrm{\AA}$ and $3.90$~$\mathrm{\AA}$ for
the LAO and STO substrates, respectively, whereas the value for
bulk LSCO is $3.82$~$\mathrm{\AA}$ \cite{Caciuffo}. In Table~1,
the in-plane (\textit{a}) and out-of-plane (\textit{c}) lattice
parameters of three different films and their tetragonal
distortion estimated as $t\!=\!2(a-c)\!/\!(a+c)$ are listed. The
third film, i.e., LSCO/PMN-PT, is epitaxially oriented but not
coherently grown. This film is used to study the response to the
reversible strain applied by the piezoelectric substrate. The
tetragonal distortion of the three films spans a wide range, from
$-2.3\%$~(LAO) to $1.8\%$~(PMN-PT) and $3.1\%$~(STO).

A drastic effect of the epitaxial strain is observed in the
electrical transport of the LSCO films. In Fig.~2, we present the
temperature dependence of the resistivity ($\rho$) of LSCO films
in different strain states. Clearly, the LSCO/PMN-PT film with
medium tensile strain displays an insulating-type behaviour.
Further, the LSCO/STO film had a Giga-ohm resistance even at
$300$~K. In contrast, the LSCO/LAO film displays metallic
conductivity at all temperatures, which is similar to the bulk
behaviour \cite{goodenough95,Lorenz,Aarbogh,Hoch}. At $5$~K , for
example, the resistivity is $\rho\simeq1.2$x$10^{-4}$ $\Omega$~cm.
We argue that this strikingly large conductivity variation among
the different films is essentially caused by the varied strain
state. The possibility of having non-connected films or
microcracks as origin of the insulating behaviour can clearly be
ruled out from the AFM images, which show fairly smooth films with
a roughness (rms) of about $1$~nm. Further, any kind of extended
defects capable to interrupt the current path would lead to strain
relaxation, which is absent for the film on STO.

Unlike the striking differences observed in the electrical
transport, the impact of biaxial strain on the magnetic ordering
is astonishingly moderate. The temperature-dependent magnetization
of the LSCO/LAO and LSCO/STO films is plotted in Fig.~3. The data
were recorded during warming in a magnetic field of $100$~mT
applied along the [100] in-plane direction for both FC and ZFC
runs. Interestingly, the magnetic ordering sets in at about
$210$~K, irrespective of the strain state. The ferromagnetic Curie
temperature ($T_{\mathrm{C}}$) estimated by extrapolating $M^{2}$
for $T\!<\!T_{\mathrm{C}}$ to $M\!=\!0$ is given in Table~1. The
$T_{\mathrm{C}}$ values are somewhat lower as compared to the
single-crystal bulk value, i.e., $225$~K \cite{Aarbogh,Lorenz},
but are in agreement with those reported by Fuchs \textit{et al.}
for low-strain films \cite{Fuchs}. The lower $T_{\mathrm{C}}$ of
our films can be then associated with finite size, thickness
effect \cite{FuchsPRB}. A pronounced branching between the ZFC and
FC curves is observed (Fig.~3), indicating a lack of long-range
ferromagnetic order and the presence of magnetic clusters in a
cluster-glass state \cite{Itoh,goodenough95}. Although there is no
significant difference in the FC curves of the two films, a higher
ZFC magnetization is observed for the LSCO/LAO film. This points
to stronger inter-cluster ferromagnetic interactions in this film.
In the framework of the DE ferromagnetism, this observation is in
line with the metallic character of the LSCO/LAO film.
\begin{figure}[!t]
\includegraphics[width=6cm]{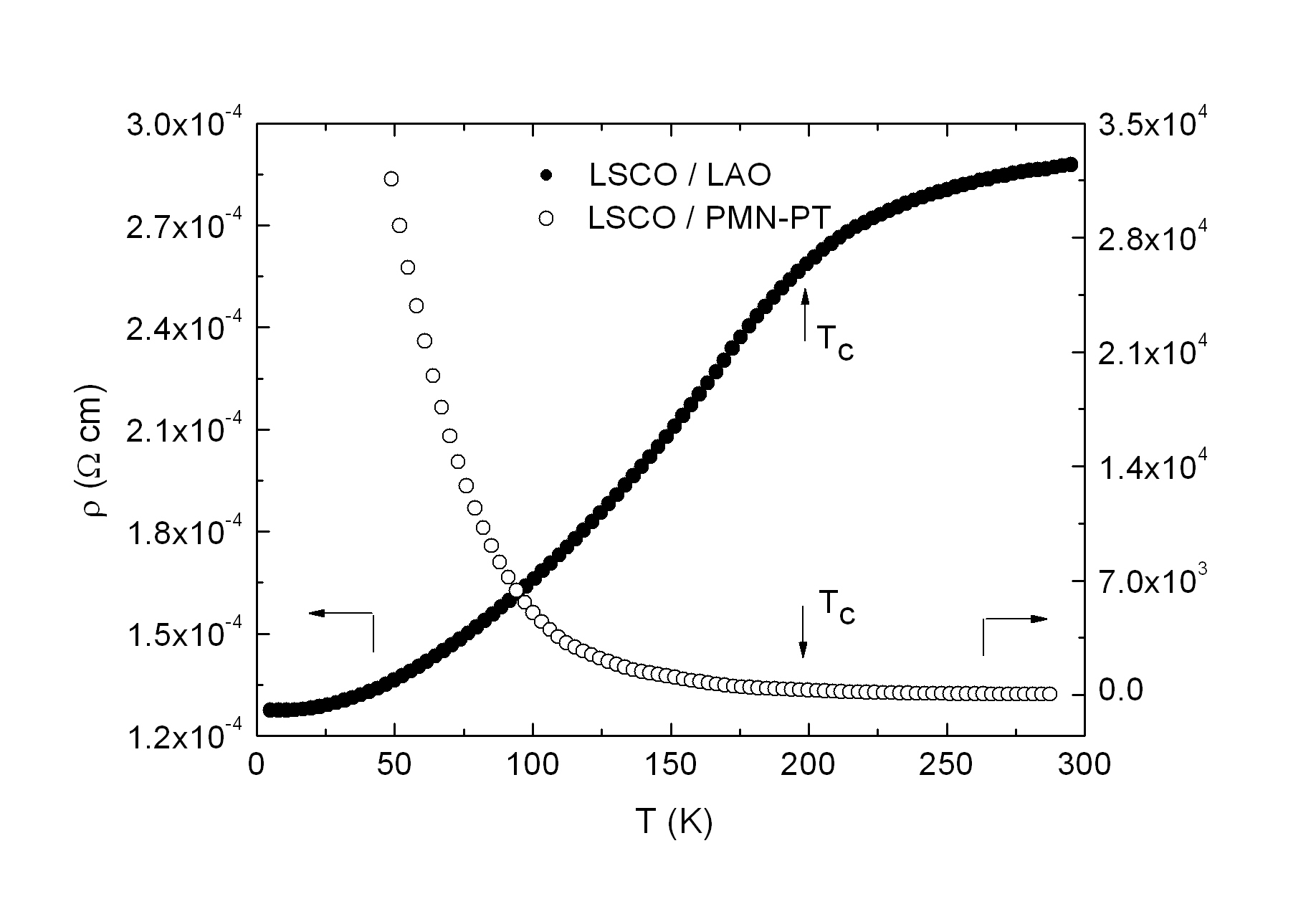}
\caption{Temperature dependence of the resistivity of LSCO~/~LAO
(left scale) and LSCO~/~PMN-PT (right scale) $60$~nm thick films.}
\end{figure}

We recorded the magnetization of a $30$~nm thick LSCO/PMN-PT film
as a function of a gradually increasing in-plane piezo-compression
which relaxes the as-grown tensile strain. A continuous increase
of $M$ is observed near $T_\mathrm{C}$, as shown in the inset of
Fig.~4b. Therefore, the increase of $M$ results from the reduced
tetragonal distortion of the unit cell. Interestingly, this is
analogue to the strain effect on $M$ in
La$_{0.7}$Sr$_{0.3}$MnO$_3$ \cite{thiele07}, where the
ferromagnetism originates from the DE mechanism. In Fig.~4b we
show the temperature dependence of the relative change of $M$
obtained by applying an electric field of $14$~kV/cm to the
substrate. This electric field produces an in-plane compression of
$0.15\%$ at $300$~K and a slightly weaker one at lower
temperatures. As for the manganites \cite{thiele07}, the strain
response peaks near $T_{\mathrm{C}}$ and vanishes for
$T\!<<\!T_\mathrm{C}$.

Transport data measured as function of the reversibly applied
piezoelectric substrate strain give direct evidence for the
extreme sensitivity to strain of the LSCO films. In Fig.~4a we
plot the resistance $R$ vs. the electric field applied to the
substrate, recorded at room temperature in a slow run, taking one
hour. It is highly hysteretic and reveals a reversible $R$
modulation by a factor of $10$. In order to exclude charging
effects the $R$ vs strain measurements were performed at various
speeds. Similar data recorded near $T_{\mathrm{C}}$ show a
modulation of $R$ by a factor of $\simeq 3$. Apart from the much
larger magnitude of the effect, the decrease of $R$ upon release
of tensile strain (i.e., reduced tetragonality of the unit cell)
agrees again with the observations for La$_{0.7}$Sr$_{0.3}$MnO$_3$
films \cite{ramesh07}. Further, it is important to note that the
strain effect we observe at $300$~K appears far above
$T_\mathrm{C}$, in the paramagnetic range.

\begin{figure}[!t]
\includegraphics[width=6cm]{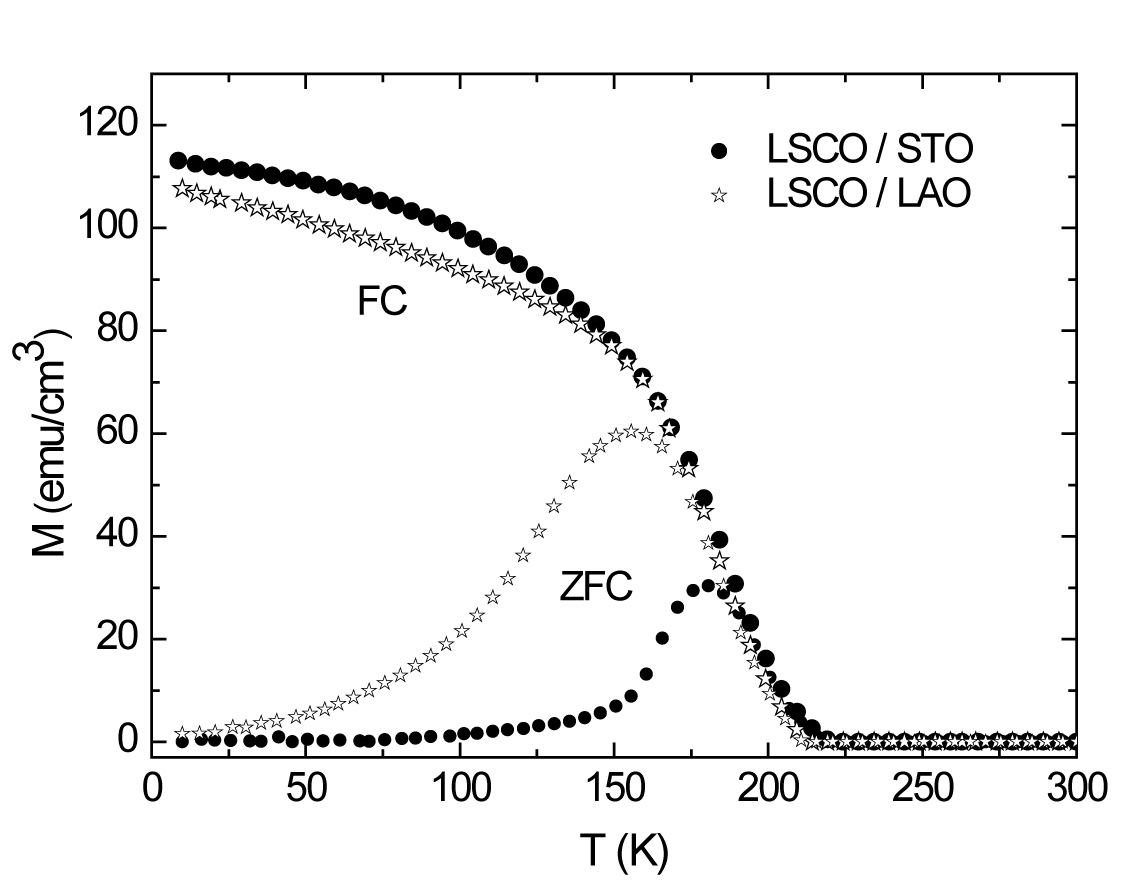}
\caption{Magnetization of LSCO films as a function of temperature
measured both in FC and ZFC mode, with a field of $100$~mT along
the [100] pseudocubic substrate direction.}
\end{figure}

Tensile strain is known to induce an insulator state in thin films
of ferromagnetic metallic manganites
\cite{kathrin_rev,YafengLu,Ziese}. For
La$_{0.7}$Sr$_{0.3}$CoO$_3$, the mechanisms already invoked for
manganites seem to play an important role, too, and will be
discussed in the next paragraph. First, the implications of a
possible spin-state transition for the Co ions is considered. The
observed pressure-induced insulating state in
La$_{0.82}$Sr$_{0.18}$CoO$_3$ is related to an enhancement of the
crystal-field splitting, driving Co$^{3+}$ ions into the LS state
and depleting the $e_g$ conduction-band
\cite{Lengsdorf,Lengsdorf2}. This mechanism is unlikely to be the
origin of the insulator state induced by tensile strain in our
films. The enlarged in-plane lattice parameter in our LSCO/STO
films is expected to reduce the splitting of the Co $t_{2g}$ and
$e_g$ levels. Additionally, the tetragonal distortion would split
the $e_g$ and $t_{2g}$ levels and result in a further reduction of
the $t_{2g}$-$e_{g}$ energy separation. Tensile strain can thus
induce a larger number of excited Co ions in the film, in contrast
to hydrostatic pressure. Still, for the doping level, $x\!=\!0.3$,
and range of temperatures, $T\!<\!T_\mathrm{C}$, we investigated,
the spin state of the Co ions seems to be rather insensitive to
strain. This is supported by the observed weak dependences of $M$,
see Fig.~3, and of the saturated magnetization on strain. Further,
$M$ increases when reducing the tensile strain, see Fig.~4b,
contrary to the expectation for a strain-induced IS or HS Co spin
state.

\begin{figure}[t]
\includegraphics[width=8cm]{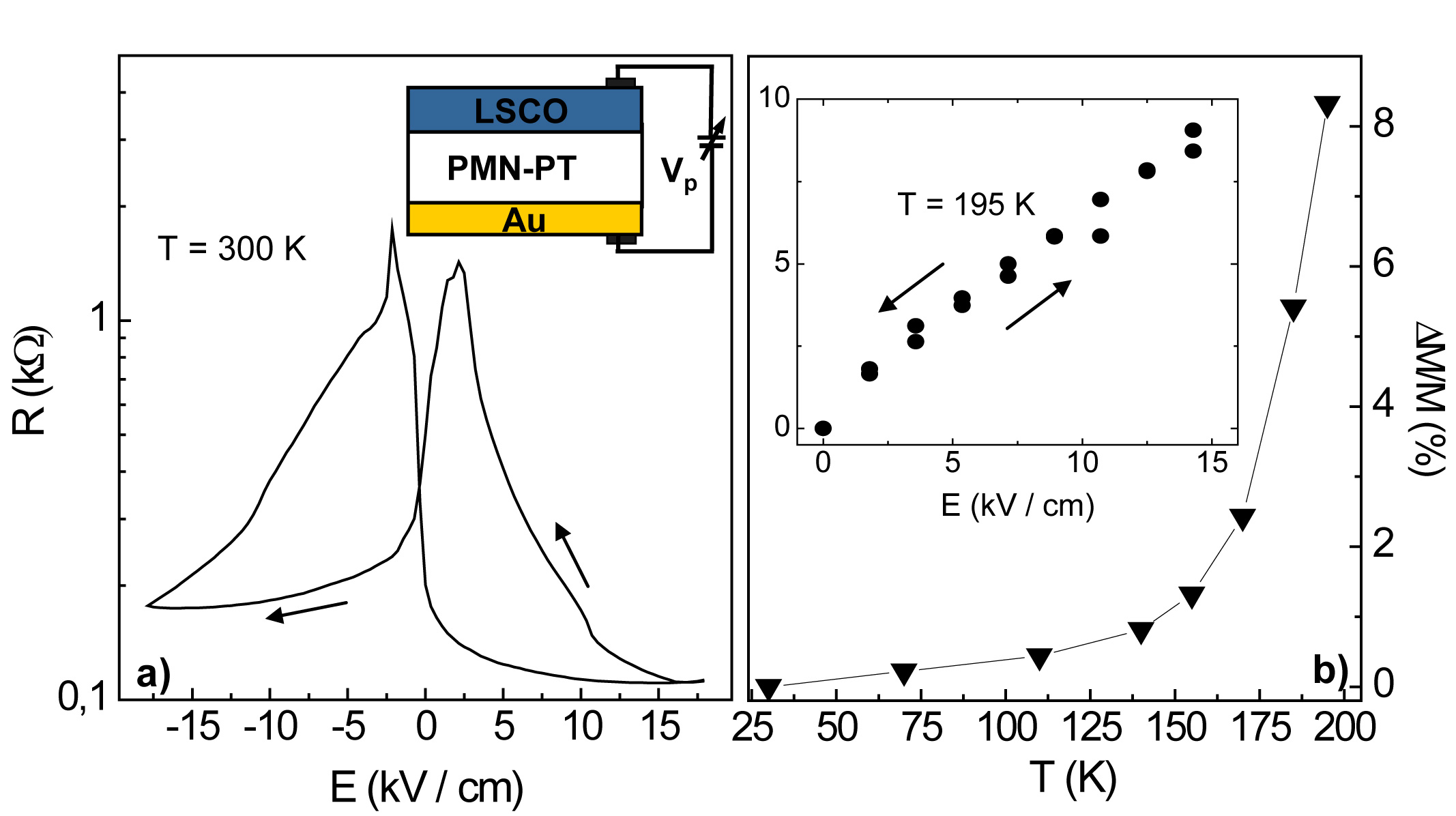}
\caption{(a) Resistance vs the electric field applied to the
substrate for a $60$~nm LSCO/PMN-PT(001) film (E//[001]). Inset
shows the schematic device structure. (b) Maximum change of $M$
calculated as $[M(E)-M(0)]/M(0)$ at various temperatures for
$E\!=\!14$ kV/cm and a $30$~nm thick film. Inset shows the change
of $M$ (in percent) with electric field, recorded at $195$~K.}
\end{figure}

Phase separation into hole-rich metallic ferromagnetic clusters
and insulating, magnetically less-ordered regions is widely
accepted for La$_{1-x}$Sr$_{x}$CoO$_3$ with $x\!<\!0.5$
\cite{Wu,Caciuffo,Hoch, Louca03,Louca06}. Magnetic field or
pressure induced cluster percolation may trigger extreme changes
of the electrical conductivity, as seen in colossal
magnetoresistive manganites \cite{tokura_rev,dagotto}. A
percolation scenario could also underlie the huge difference in
conductivity for our differently strained films as well as the
extremely large direct strain effect (Fig.~4a). The observation of
the latter at $300$~K goes beyond the characteristics of the CMR
manganites because it occurs in the paramagnetic state, indicating
the existence of an electronic cluster state at
$T>>T_{\mathrm{C}}$. The magnetization data (Fig.~3) resemble the
characteristics of a cluster state as typically observed in the
bulk. Furthermore, we found a strong increase of the coercive
field for the LSCO/STO ($879$~mT at $10$~K) as compared to
LSCO/LAO ($440$~mT at $10$~K). This points to weaker inter-cluster
coupling in the LSCO/STO film, which is in line with the
insulating state. A phase separation scenario could also explain
the strikingly weak change of $T_\mathrm{C}$ with strain. It seems
that $T_\mathrm{C}$ is mainly determined by the intra-cluster
ordering. At this point it is worth noting the rather weak
dependence of $T_\mathrm{C}$ with doping, i.e.,
$220$~K$<\!T_\mathrm{C}\!<\!270$~K, in bulk
La$_{1-x}$Sr$_{x}$CoO$_3$ with $x$ varying between $0.2$ and $0.9$
\cite{goodenough95,Croft}, which may as well be related to a
magnetic phase separation.

At the microscopic level, the strain effect in ferromagnetic
manganites has been essentially attributed to changes in the
ferromagnetic DE and the Jahn-Teller splitting of the $e_{g}$
levels \cite{tokura_rev,dagotto,kathrin_rev}. In nearly cubic
compounds, the strain-induced tetragonal distortion of the unit
cell produces (i) a suppression of the DE mechanism and (ii)
static Jahn-Teller type distortions favoring the occupation of the
in-plane $e_g$ $d_{x^{2}-y^{2}}$ orbitals
\cite{YafengLu,Ziese,Fang}. Both effects lead to electron
localization. A similar interplay between the DE and
strain-induced Jahn-Teller/orbital ordering may be active in
cobaltites and could be responsible for the observed
strain-induced insulator state in La$_{0.7}$Sr$_{0.3}$CoO$_3$.
This idea is supported by the fact that the same dependences were
found with strain for both magnetization and resistance in the
direct-strain experiments on La$_{0.7}$Sr$_{0.3}$MnO$_3$ and
La$_{0.7}$Sr$_{0.3}$CoO$_3$: reduced tetragonality leads to larger
magnetization and lowers the resistivity. On the other hand, the
electronic structure and, thus, transport properties in cobaltites
may be even more sensitive to structural changes as compared to
the manganites. Reliable \textit{ab initio} calculations are
strongly desirable to clarify these issues. Additionally,
cobaltites may offer access to strain-induced control of
electronic properties in the paramagnetic range at temperatures
above $300$~K.

Finally, it is ruled out that compositional differences among the
films cause the different conductivity. Tensile strain could
result in oxygen deficiency and/or Sr enrichment. In bulk, oxygen
vacancies up to $5\%$ lead to an increase of $R$ by less than a
factor of $5$ \cite{Hsu}. Higher Sr doping would drive the films
metallic according to the bulk phase diagram. Considering also the
huge direct resistive strain effect, we argue that composition is
unlikely to be responsible for the observed variation of the
conductivity.

To summarize, the effects of both static and dynamic biaxial
strain on electrical transport and magnetization of
La$_{0.7}$Sr$_{0.3}$CoO$_3$ films have been investigated.
Depending on the strain state, we observed extreme changes in the
conductivity. In particular, tensile strain above $1\%$ is found
to induce a transition to an insulator state. Reversible strain of
$0.15\%$ triggered huge resistance modulations, including a change
by a factor of $10$ in the paramagnetic regime at $300$~K. Whereas
epitaxial strain may generally be applied to control the spin
state in cobaltite films due to their small crystal field
splitting, the weak dependence of the magnetization on strain
indicates only a minor modification of the spin state in
La$_{0.7}$Sr$_{0.3}$CoO$_3$ below $T_\mathrm{C}$. We interpret the
changes observed in the transport properties in terms of a
strain-induced splitting of the Co $e_{g}$ levels and reduced
double exchange, combined with a percolation-type conduction in an
electronic cluster state.

We would like to thank D. Khomskii, M. Richter, L. Hozoi, V.
Kataev, K.~H. M\"{u}ller, H. Tjeng and L. Eng for fruitful
discussions and C. Richter for the help with electrical
measurements. This work was supported by Deutsche
Forschungsgemeinschaft, FOR $520$.

\end{document}